\documentclass{aa}
\usepackage{graphicx}
\begin{document}
   \title{Carbon Monoxide in type\,II supernovae\thanks{Based on observations
   collected at the European Southern observatory, Chile in time
   allocated to proposals 62.H-0708 and 65.H.-0462}}


   \author{J. Spyromilio, 
          B. Leibundgut 
          and
          R. Gilmozzi 
          }

   \offprints{J. Spyromilio}

   \institute{European Southern Observatory,
              Karl-Schwarzschild-Strasse 2, Garching, D-85748, Germany \\
              \email{jspyromi@eso.org, bleibundgut@eso.org, rgilmozz@eso.org}
             }

   \date{Received ; accepted }

   \abstract{Infrared spectra of two type\,II supernovae 6 months
   after explosion are presented. The spectra exhibit a strong
   similarity to the observations of SN\,1987A and other type\,II SNe
   at comparable epochs. The continuum can be fitted with a cool black
   body and the hydrogen lines have emissivities that are
   approximately those of a Case B recombination spectrum. The data
   extend far enough into the thermal region to detect emission by the
   first overtone of carbon monoxide. The molecular emission is
   modeled and compared with that in the spectra of SN\,1987A. It is
   found that the flux in the CO first overtone is comparable to that
   found in SN\,1987A. We argue that Carbon Monoxide forms in the
   ejecta of all type\,II SNe during the first year after explosion.
   }

\titlerunning{CO in type\,II SNe}

   \maketitle
%

\section{Introduction}

The presence of molecular emission in the spectra of supernovae while
not new remains rarely detected. First detection of carbon monoxide in
a supernova was in the spectra of SN\,1987A (Catchpole \& Glass
\cite{catchpole}, McGregor \& Hyland \cite{mcgregor}, Ames Research
Center \cite{ames}, Oliva et al. \cite{oliva}, Spyromilio et
al. \cite{spyromilio1}).  In SN\,1987A observations of the first
overtone of CO at 2.3-$\mu$m were complemented by observations at
4.6-$\mu$m of the fundamental bands.  In addition to carbon monoxide,
the spectra of SN\,1987A revealed bands of SiO (Meikle et
al. \cite{meikle2}; Roche et al. \cite{roche}) and CS (Meikle et
al. \cite{meikle2}). The presence of H$_3^+$ is also claimed in the
spectra of SN\,1987A (Miller et al. \cite{miller}) although
alternative atomic species can also explain the identified features.
Spyromilio \& Leibundgut (\cite{spyromilio2}) reported the detection
of CO first overtone emission in the spectra of the type\,II supernova
1995ad while Gerardy et al. (\cite{gerardy}) and Fassia et
al. (\cite{fassia2}) reported the presence of  CO in SN\,1998S.

The formation of molecules in the ejecta of supernovae, even CO - the
most stable of diatomic molecules with a dissociation energy of
11.09\,eV (Douglas \& M{\o}ller \cite{douglas}), is not trivial. The
high UV field due to recombinations and the energetic electrons from
the radioactive decays of $^{56}$Ni and its daughter $^{56}$Co create
an inhospitable environment with multiple dissociation paths (see
Lepp, Dalgarno \& McCray \cite{lepp}). The presence therefore of
molecules in the ejecta imply a particular distribution of the
material within the ejecta and place constraints on the degree of
mixing possible. These effects have only been analyzed in detail in
the well observed SN\,1987A. The degree to which that object is
representative of the entire class of type\,II SNe, which will almost
by definition not be observed with the same accuracy, is of some
importance. 

Molecules also affect the degree of cooling of the ejecta. The
partition function of a diatomic molecule such as CO is enormous
compared to that of even a heavy ion such as Fe$^+$. Moreover, the
lower rotational levels of the fundamental, first and second overtones
of CO require low energies for their excitations allowing them to
continue cooling the ejecta even as the temperature drops. 

Gerardy et al. (\cite{gerardy}) and Fassia et al. (\cite{fassia2})
also report the detection of emission from dust in SN\,1998S and
Fassia et al. argue that the cooling by CO may lead to conditions that
favour the formation of dust. 

Here we report on observations of two more type\,II SNe in the near
infrared which exhibit emission by carbon monoxide. In sect.\,2 we
report on the observations. The data are discussed and compared with those
of SN\,1987A and theoretical models in sect.\,3.

\section{Observations}

SN\,1998dl in NGC1084 (recession velocity 1411\,kms$^{-1}$) 
was discovered by King et al. (\cite{king}), on
1998 August 20, from observations made at the 0.8-m Katzman Automatic
Imaging Telescope (KAIT). It was subsequently found to have been present in
images taken on 1998, August 2 and August 9. Filippenko
(\cite{filippenko}) detected hydrogen Balmer lines in early optical
spectra classifying the object as a type\,II supernova.

SN\,1998dl was observed at the ESO 3.58-m New Technology Telescope on
La Silla, on 1999 January 1, at an age of approximately 150 days.  The
near-infrared camera and spectrograph SofI (Moorwood et
al. \cite{moorwood}) was used for these observations. The red
low-resolution grism was used for the $H$ and $K$ band spectra. While
the grism does cover the region between 1.8 and 1.92\,$\mu$m the
atmospheric transparency is very low and no useful data are available
in this region. A 1 arcsecond slit was used which gives a resolution
of $\lambda$/$\Delta \lambda$ of 650 and a wavelength coverage from
1.5\,$\mu$m to 2.5-$\mu$m. The spectroscopic standard BS\,2290 (Allen
\& Cragg \cite{allen}) was used to remove atmospheric features and to
correct for the spectral response of the instrument. The
conditions were not photometric and therefore the absolute fluxing of
the spectrum is arbitrary. The slit projects to 3 pixels on the
detector oversampling the data.  Narrow features, of order 3 pixels
(e.g. the particularly prominent feature at 2\,$\mu$m), in the spectra
are not real but result from incomplete corrections for atmospheric
features.

Supernova 1999em in NGC 1637 (recession velocity 717\,kms$^{-1}$) was
discovered by Li (\cite{li}) from observations made, on 1999 October
29, at the KAIT. Jha et al. (\cite{jha}) and Deng et al. (\cite{deng})
classified the object as a type\,II supernova based on spectroscopic
observations.

SN\,1999em was observed, on 2000 April 24, at an age of approximately
170 days with the same instrumental configuration as 1998dl. The
spectroscopic standard used in this case was HIP 27855 (Perryman et al.
\cite{perryman}). The accurate removal of atmospheric features
requires that the standard and the target are observed using the same
resolving power. The usage of a narrow slit in both observations of
the standard and the target implies that a separate photometric
calibration of the data needs to be made to recover absolute
fluxes. The imaging mode of SofI was used for this purpose and the
NICMOS standard S121-E (Persson et al. \cite{persson}) was used for the
calibration. The derived $K_s$ magnitude of the supernova was
15.1. The mean flux of the observed spectrum was corrected to the
this magnitude.

\section{Discussion}

\begin{figure}
\centering
\includegraphics[angle=-90,width=9cm]{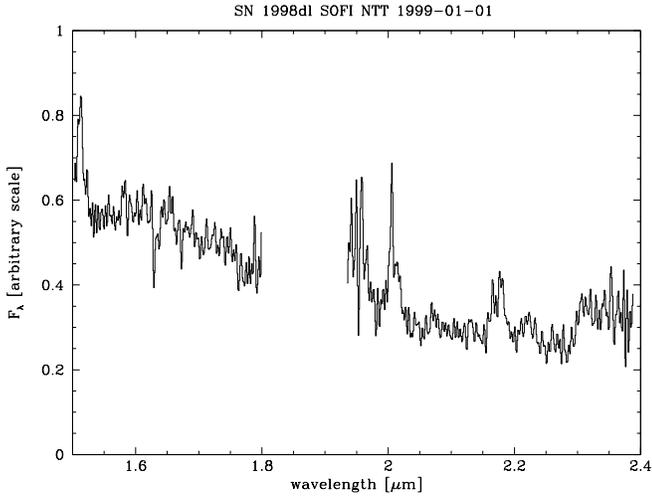}
 \caption{Spectrum of 1998dl.}
\label{98dlraw}
\end{figure}

\begin{figure}
\centering
\includegraphics[angle=-90,width=9cm]{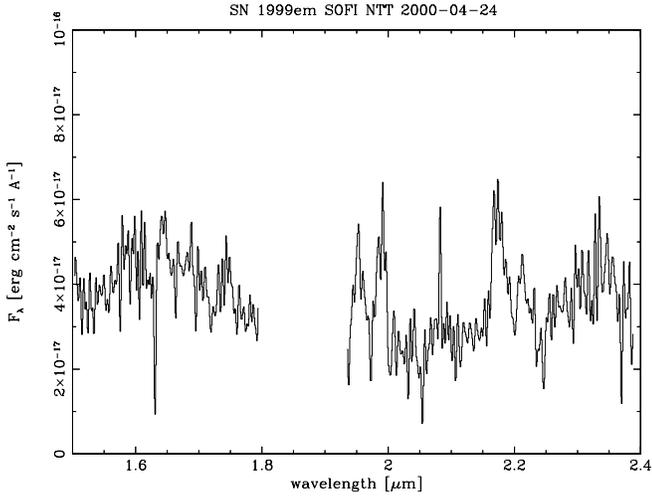}
 \caption{Spectrum of 1999em.}
\label{99emraw}
\end{figure}

The spectra of SN\,1998dl and SN\,1999em are displayed in
Figs.\,\ref{98dlraw} and \ref{99emraw}, respectively. The data are
shown at their observed wavelengths without correction for the
recession velocity of the parent galaxies.  The spectra are
characterized by a strong continuum upon which emission lines are
superimposed. The strong Brackett\,$\gamma$ line (rest wavelength
2.1655\,$\mu$m) is evident. In the SN\,1999em data the shorter
wavelength transitions of the Brackett series can also be
seen. Long-wards of the Br$\gamma$ line, at 2.21\,$\mu$m, a broad
feature identified by Meikle et al. (\cite{meikle1}) in the spectra of
SN\,1987A as emission by the 2.207\,$\mu$m Na{\small I}
4s$^2$S--4p$^2$P$^o$ multiplet is evident\footnote{In Meikle et
al. (\cite{meikle1}) the lower level of this multiplet in table\,3 is
incorrectly labeled as 4s$^2$D}. Starting at 2.3\,$\mu$m and extending
to the end of our spectral converage a broad
feature is evident in the data. This broad feature we attribute to carbon
monoxide first overtone transitions ($\Delta v = 2$). The much weaker
second overtone ($\Delta v = 3$) is expected to start at
1.5\,$\mu$m. Its contribution is expected to be only 1.4\% of the
strength of the first overtone (Bouanich \& Brodbeck \cite{bouanich})
and is therefore not detectable in our spectra. In much higher signal
to noise data as would possibly be obtained at the VLT with ISAAC it
might be possible to detect such emission. The fundamental ($\Delta v
= 1$) band at 4.6\,$\mu$m is very much stronger than the first
overtone observed here but also in an extremely unfavourable
atmospheric window. This fundamental band, as mentioned in the
introduction, almost dominates the cooling of the ejecta for some time
and observation would be of great interest. Unfortunately they are
probably limited to a combination of the very biggest telescopes and
the very nearest supernovae.

A strong emission feature is observed at 2\,$\mu$m. As noted above the
narrow spike superimposed on the broad emission is instrumental. In
SN\,1987A, a broad feature at the same wavelength was identified by
Meikle et al. as emission by the Ca{\small I} 4p$^3$P$^o$--3d$^3$D
multiplet and the [Fe{\small I}] a$^5$F$_5$--a$^3$F$_4$
multiplet. Assuming a contribution to the 2\,$\mu$m by [Fe{\small I}]
in the spectrum of SN\,1998dl, the strong line at 1.51\,$\mu$m could be
identified as emission by the [Fe{\small I}] a$^5$D$-2$--a$^5$F$_4$
multiplet. Alternatively Mg{\small I} also has a strong transition at
1.5031\,$\mu$m.  The lower recession velocity of the parent galaxy of
SN\,1999em prevents us from detecting the same line in that spectrum.

At 2.06\,$\mu$m both in SN\,1998dl and SN\,1999em a small absorption
is present. This we identify as the P-Cygni trough of the He{\small I}
2s$^1$S--2p$^1$P$^o$ 2.058\,$\mu$m transition. We note that this
region is one of poor atmospheric transmission with a strong telluric
CO$_2$ feature at 2.06\,$\mu$m. While we believe that we have
correctly compensated for this it is possible that the feature we
identify as due to He{\small I} is due to poor cancellation by of 
aforementioned atmospheric feature. He{\small I} is expected to be
very weak but its presence would be significant as highly excited He is a
tracer for the presence of energetic electrons from the radioactive
decay (Graham \cite{graham}; Fassia \& Meikle \cite{fassia}). The
singlet transition observed here would need to develop sufficient
optical depth to exhibit a P-Cygni trough. While transitions from
2s$^1$S to the ground state of He{\small I} are forbidden,
depopulation of this energy level can occur via two photon decay with
a low transition probability (A=51s$^{-1}$). Given the age of the
supernovae at the time of observation even this process would have
de-populated the ground state of the 2.058\,$\mu$m
line. Recombinations are the obvious source of electrons which, as the
transitions from the $^1$P$^o$ states to the ground are expected to be
saturated, will naturally cascade down to the 2s$^1$S level. Given the
high ionization potential of He (24.587\,eV) the presence of the
2.058\,$\mu$m transition suggests that the excitation by energetic
electrons argued by Graham and Fassia \& Meikle is occuring also in
these two SNe.

\begin{figure}
\centering
\includegraphics[angle=-90,width=9cm]{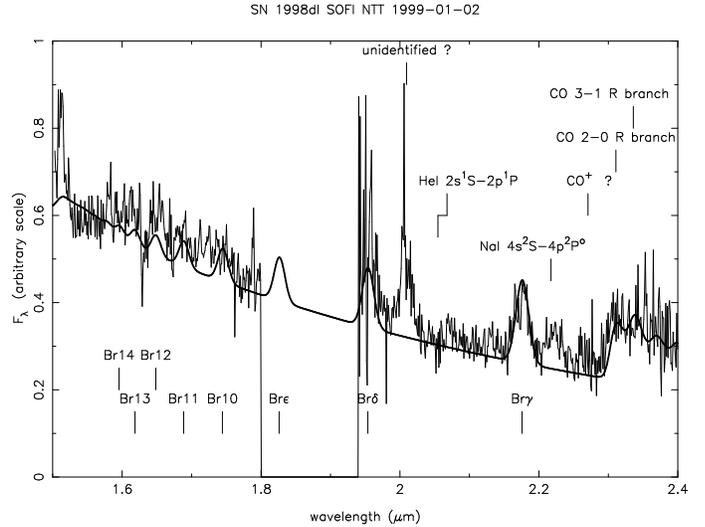}
 \caption{Spectrum of 1998dl and model. Features discussed in the text
are identified in the figure.}
\label{98dlfit}
\end{figure}

\begin{figure}
\centering
\includegraphics[angle=-90,width=9cm]{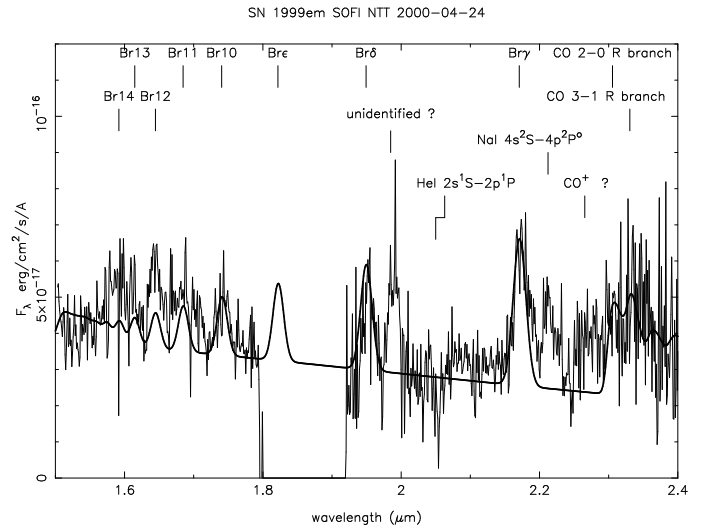}
 \caption{Spectrum of 1999em and model.  Features discussed in the text
are identified in the figure.}
\label{99emfit}
\end{figure}

We have used simple line models to make fits to the data. These fits are
shown in Figs.\,\ref{98dlfit} and \ref{99emfit}. The ingredients of the
models are described below. 

The continuum in the models is a black-body with a temperature of
3300\,K for SN\,1998dl and 2000\,K for SN\,1999em. The flux observed
in the spectrum of SN\,1999em at this temperature corresponds to a
surface of last scattering expanding at 500\,kms$^{-1}$ for 170 days
at a distance of 7.8\,Mpc (Sohn \& Davidge \cite{sohn}).

For the carbon monoxide a simple LTE (Boltzmann distribution) model of
the populations of the rotational-vibrational energy levels in the
ground state electronic level $^1\Sigma$ of $^{12}$C$^{16}$O (CO) was
used. This model has been described before in Spyromilio et
al. (\cite{spyromilio1}). Briefly, the energy levels for a given
rotational level $J$ and a given vibrational level $v$ can be
determined using:

\begin{equation}
\begin{array}{l}
E = \omega(v + \frac{1}{2}) - \omega x (v + \frac{1}{2})^2 
+ \omega y (v+ \frac{1}{2})^3 \\
 + B_vJ(J+1) - D_vJ^2(J+1)^2 
\end{array}
\end{equation}

with $\omega , \omega x, \omega y, B_v$ and $D_v$ from Young
(\cite{young}). The A values are given by 

\begin{equation}
A_{vv'} = \frac{64 \pi^4 \nu^3}{3 h g_J}(TM)^2
\end{equation}

where $\nu$ is the transition frequency, $g_J = 2J +1$ and TM is the
transition matrix from vibrational level $v$ to $v'$. TM for $\Delta
J$ of $\pm$1 is given by Chackerian \& Tipping (\cite{chackerian})
The lowest 15 vibrational levels were included and within each 110
rotational level populations were calculated. 

As has been shown in the spectra of SN\,1987A (Spyromilio
et al. \cite{spyromilio1}) the CO band heads are sensitive to
the excitation temperature.  
Higher temperature in the CO model increase the contribution of higher
vibrational pairs and increase the emission long-wards of the $v=2$ to
$v'=0$ band-head. The quality of the data is not good enough to
reliably distinguish between small changes in the temperature. The
SN\,1998dl data indicate a slightly higher temperature than the
SN\,1999em. The fits use 2500\,K for SN\,1998dl and 2000\,K for
SN\,1999em. The uncertainty in these values is of order a few hundred
degrees K. The difference in the temperatures leads to a significantly
different appearance of the first overtone emission. The 2000\,K
temperature for the SN\,1999em spectrum is based on the assumption
that the drop in flux long-wards of 2.35\,$\mu$m is real. The
SN\,1998dl spectra appear to show a much smoother decline towards longer
wavelengths.  

The velocity at which the CO is expanding can be used to constrain the
mass of the progenitor (Gerardy et al. \cite{gerardy}). In SN\,1998S
Gerardy et al. derive a high expansion velocity ($v \approx
3000$\,kms$^{-1}$) based on the smooth rise of the 2-0 R-branch band
head and a progenitor mass in excess of 25\,M$_{\odot}$. Fassia et
al. (\cite{fassia2}) do not detect such a steep rise and argue for a
lower velocity. In addition their modeling of the ejecta favours a
lower mass progenitor. We argue below in favour of the low range of
velocities although we, like Gerardy et al., have a smooth rise of the
2-0 R-branch band head. 

In the spectra of SN\,1987A a rigorous fitting of the expansion
velocity could be made (Spyromilio et al. \cite{spyromilio1}) based on
the appearance of the R-branch emission from each  pair of 
energy states.  Both Gerardy et al. (\cite{gerardy}) and Fassia et
al. (\cite{fassia2}) use the lack of clear separation of the R-branch
band heads to place lower limits on the expansion velocity in the
observations of SN\,1998S. This same effect can be used by us to place
a weak lower limit of 1000\,kms$^{-1}$ on the expansion velocity.  

\begin{figure}
\centering
\includegraphics[angle=-90,width=9cm]{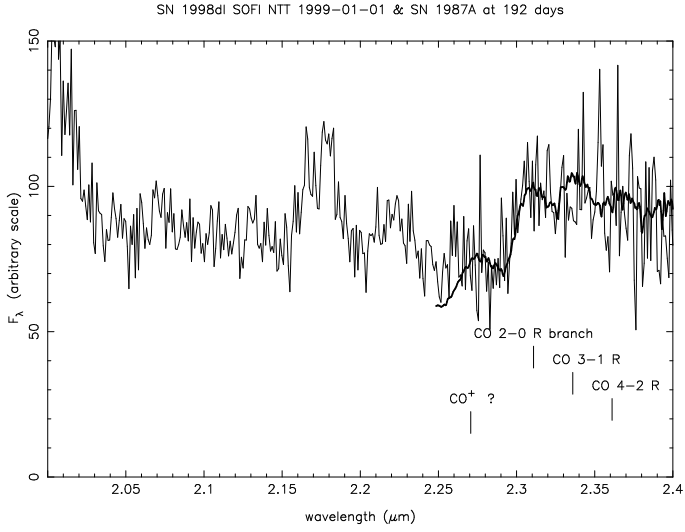}
 \caption{Spectrum of 1998dl in the $K$ band with 
spectrum of SN1987A superimposed.}
\label{98dl87A}
\end{figure}

\begin{figure}
\centering
\includegraphics[angle=-90,width=9cm]{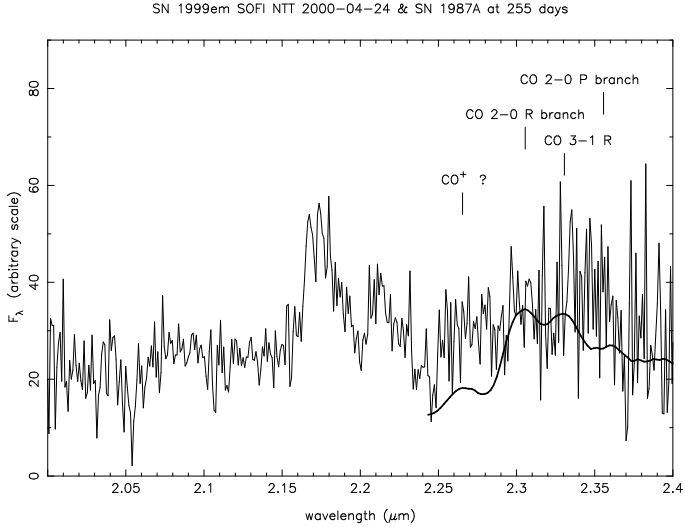}
 \caption{Spectrum of 1999em in the $K$ band with 
spectrum of SN1987A superimposed. Note that the feature marked as
CO$^+$ is much weaker in 87A than in 99em.}
\label{99em87A}
\end{figure}

The comparison of SN\,1998S with SN\,1987A (see Gerardy et
al. Fig.\,6) shows that another line is possibly blended with the
CO band head at 2.28\,$\mu$m. This feature, which at the recession
velocity of SN\,1987A is observed at 2.26\,$\mu$m, is discussed later
but remains unidentified. The 2.26\,$\mu$m feature is very strong in
the spectra of SN\,1999em and weakly present in our data on
SN\,1998dl. The same feature was present in the spectra of SN\,1995ad
(Spyromilio \& Leibundgut \cite{spyromilio2}). We cannot account for
the lack of detection of this in the data of Fassia et
al. (\cite{fassia2}). The blending of this feature into the band head
implies that models that do not take this into account will
underestimate the true sharpness of the band head rise and derive
higher than expected expansion velocities. The best estimate of the
velocity of the CO, we believe, comes from the comparison of the
observed spectra with those of SN\,1987A (see Figs.\,\ref{98dl87A}
and \ref{99em87A}). The good matching of the rise of the 2-0 band head
suggests that the velocity observed in SN\,1987A
(1800--2000\,kms$^{-1}$; Spyromilio et al. \cite{spyromilio1}) is
representative of the velocity of the CO in SN\,1998dl and
SN\,1999em. Clearly in the absence of the blended line we would agree
with Gerardy et al. (\cite{gerardy}) and also derive a much higher
expansion velocity for the CO. Each CO line in our models was
convolved with a Gaussian line profile of FWHM 2000\,kms$^{-1}$.

The Brackett series line strengths in the model assume a case-B
recombination spectrum at a temperature of 3000\,K and electron
density of 10$^5$ (Hummer \& Storey \cite{hummer}). The Br$\gamma$ 
and $9 \rightarrow 4$ transitions are well fit using
these parameters. The Br$\delta$ line lies at the very edge of the
atmospheric window and the cancellation of the atmospheric features
there is poor. While the norm would be not to display the spectra in
this region due to the large uncertainties in the data we include them
here. The data show that the spectra are consistent with the model. We
do not derive any further information from the region around
1.92\,$mu$m. Unfortunately the Br$\epsilon$ transition lies in
the gap between the two atmospheric windows and is not observable from
the ground. The FWHM of the Gaussian line profile computed was
3000\,kms$^{-1}$. As has been the case in the other SNe in which CO was
detected, the velocity of the hydrogen lines is higher than that of the
CO.

The individual lines identified earlier as Na{\small I},
Ca{\small I} and [Fe{\small I}] have not been included in the displayed
fits as the individual contributions of the species would be arbitrary.

As already discussed with respect to the expansion velocity, the
observations presented here can be compared with the observations of
SN\,1987A (data from Meikle et al. \cite{meikle1}).
Fig.\,\ref{98dl87A} shows the data from SN\,1987A taken at an age of
192 days superimposed on the spectra of SN\,1998dl. The SN\,1987A data
have been scaled to match the flux in the CO feature and red-shifted to
the recession velocity of NGC1084. The continuum underlying the
SN\,1987A data has been artificially adjusted to match the average
continuum level in the SN\,1998dl data. The SN\,1998dl continuum is
much bluer than that of SN\,1987A at the epoch at which the
comparison data are taken. The poor correlation at 2.25\,$\mu$m is
partially due to this effect and partially due to the difference in
the strength of the 2.26\,$\mu$m feature. Given the quality of the SN\,1998dl
data, this comparison, adds plausibility to the identification of CO
in the spectra. Also as mentioned above it allows a more accurate
determination of the velocity structure of CO than can be achieved
from the models. A similar comparison is made in Fig.\,\ref{99em87A}
for SN\,1999em using SN\,1987A data obtained at an age of 255 days.

The flux within the CO emission band for SN\,1999em is 5.5$\times$10$^{-11}$
ergs/s/cm$^2$ which at the distance of NGC 1637 (7.8\,Mpc; Sohn \&
Davidge \cite{sohn}) is 20\% lower than that of SN\,1987A at a comparable
epoch. 

The mass that corresponds to this emission depends critically on the
emission model. Our LTE model gives masses of order
10$^{-4}$\,M$_{\odot}$. Liu \& Dalgarno \cite{liu} predict a much higher
mass ($\sim$10$^{-3}$\,M$_{\odot}$) for the same emission at this
age and argue that the conditions within the ejecta favour the
formation of this amount of CO. In fact it is expected that
significantly more CO formed during the youth of the supernova
($\approx 100$ days after explosion) but has been destroyed. Gearhart,
Wheeler \& Swartz \cite{gearhart} favour the formation of much less CO
than Liu \& Dalgarno and more in line with our mass estimates. Both
Gerardy et al. (\cite{gerardy}) and Fassia et al. (\cite{fassia2})
employ non-LTE and optically thick models to estimate the mass of CO
emitting and derive high masses. These models base their input on
assumptions and parameters that cannot be derived from our limited
observations and even in the case of SN\,1998S are based on assumed
filling factors. Moreover Gerardy et al. argue that their comparisons
are more {\em a proof of principle} rather than true models.  We
restrict ourselves therefore to our comparison with
SN\,1987A. Complications due to clumping and asphericity
are beyond the scope of this work to address but it is worth noting
that polarization has been detected in the spectra of SN\,1999em
(Leonard, Filippenko \& Chornock \cite{leonard}; Wang et
al. \cite{wang}).

Sollerman et al. (\cite{sollerman}) detected no sign of dust in
spectroscopic observations of SN\,1999em obtained at an age of 450
days. While the formation of molecules may provide the
cold sites for dust formation in the ejecta of type\,II supernovae and
therefore may be a necessary precondition, it is clearly not
sufficient.

The feature at 2.26-$\mu$m short-wards of the CO $v = 2 \rightarrow 0$
band head remains unidentified. The formation of sufficient CO$^+$,
whose first overtone $v = 2 \rightarrow 0$ band head coincides with
this feature, is excluded by both Gearhart et al. and Liu \& Dalgarno.
Note, however, that Petuchowski et al. \cite{petuchowski} have argued
that CO$^+$ could form in sufficient quantities. The feature was
reported by and identified as CO$^+$ by Spyromilio et
al. (\cite{spyromilio1}) and reported by Spyromilio \& Leibundgut in
the spectra of SN\,1995ad. It is also observed by Gerardy et
al. (\cite{gerardy}) in SN\,1998S. Fassia et al. (\cite{fassia2}) did
not detect the emission in their spectra of SN\,1998S. While
theoretical models fail to predict the necessary amounts of CO$^+$ the
identification of this feature remains insecure.

\section{Conclusions}

The spectra presented in this paper combined with the observations
presented by Spyromilio \& Leibundgut (\cite{spyromilio2}), Gerardy et
al. (\cite{gerardy}) and Fassia et al. (\cite{fassia2}) suggest that
the formation of CO in the ejecta of type II supernovae is ubiquitous
at an age of 3 to 6 months. 

The mass of CO present in the ejecta is difficult to estimate
accurately. Our LTE models while simplistic rely on few assumptions
with respect to the distribution and excitation of the CO. While for
well studied supernovae such as SN\,1987A the much more detailed
models of Lepp, Dalgarno and McCray (\cite{lepp}) are well justified, 
it is arguable whether the input parameters for such models are well
enough constrained in the case of the objects described here. We
therefore argue that since the absolute fluxes argee for SN\,1987A and
SN\,1999em so, to first approximation, will the emitting masses. 

The expansion velocity of the CO is not accurately determined. 
By comparison with SN\,1987A we argue that velocities around
2000\,kms$^{-1}$ are consistent with our data. We also argue that the
blending of the ubiquitous (in our data) feature at 2.26\,$\mu$m with
the 2-0 R branch band head has to be taken into account when using
this profile for velocity determinations. 

We note that comparing the spectra of the four other supernovae with
CO emission with those of SN\,1987A it is consistently the case that
the spectral signature of the emission resembles that of SN\,1987A at a
later age than that at which the other SNe are observed.

The other aspects of our data indicate that the spectra of type\,II SNe
in the near IR are very similar and that recombination continues to
play a major role even 5 to 6 months after explosion. 

A better understanding of the physical conditions in the ejecta of
supernovae can be achieved with higher quality infrared data as would be
obtained with ISAAC at the VLT. Very high quality data may even
constrain the isotopic ratios of carbon and oxygen as the wavelengths
of the band heads to shift depending on the exact isotopes. 

\begin{acknowledgements}

We thank the staff at the NTT for their excellent support over many
years of observing. We thank David Silva for outNEDing NED.  This
research has made use of the NASA/IPAC Extragalactic Database (NED)
which is operated by the Jet Propulsion Laboratory, California
Institute of Technology, under contract with the National Aeronautics
and Space Administration. 

\end{acknowledgements}

\end{document}